\documentclass[12pt]{iopart}
\usepackage{graphicx,iopams}

\begin{document}

\title{Bogoliubov-de Gennes study of trapped spin-imbalanced unitary Fermi
gases}

\date{\today}

\author{L. O. Baksmaty$^{1}$, Hong Lu$^{1}$, C. J. Bolech$^{2,1}$ and
Han Pu$^{1}$ }

\address{$^{1}$Department of Physics and Astronomy and Rice Quantum Institute,
Rice University, Houston, TX 77005, USA}

\address{$^{2}$Department of Physics, University of Cincinnati, Cincinnati,
OH 45221, USA}
\begin{abstract}
It is quite common that several different phases exist simultaneously
in a system of trapped quantum gases of ultra-cold atoms. One such
example is the strongly-interacting Fermi gas with two imbalanced
spin species, which has received a great amount of attention due to
the possible presence of exotic superfluid phases. By employing novel
numerical techniques and algorithms, we self-consistently solve the
Bogoliubov de-Gennes equations, which describe Fermi superfluids in
the mean-field framework. From this study, we investigate the novel
phases of spin-imbalanced Fermi gases and examine the validity of
the local density approximation (LDA), which is often invoked in the
extraction of bulk properties from experimental measurements within
trapped systems. We show how the validity of the LDA is affected by
the trapping geometry, number of atoms and spin imbalance. 
\end{abstract}

\pacs{03.75.Ss, 71.10.Ca, 37.10.Gh}

\maketitle
\noindent \textit{Keywords\/}: Fermi gas, Fulde-Ferrel-Larkin-Ovchinnikov
state

\section{Introduction}

Interest in spin-imbalanced Fermi superfluids dates back over a half
century to the appearance of the Bardeen-Cooper-Schrieffer (BCS) theory.
Several pioneers theoretically considered the fate of a superconductor
in the presence of magnetic impurities which could disrupt the spin
balance \cite{clogston,chandra,FF_original,LO_original} prescribed
by BCS. In the ensuing decades the BCS picture has emerged as an important
paradigm in many branches of physics \cite{casalbuoni}, such as nuclear
physics, quantum chromodynamics, and ultra-cold atoms. The ultra-cold
atoms, in particular, provide a highly controllable clean system to
investigate the effect of spin imbalance on the nature of fermionic
pairing. Over the past few years, experiments on polarized Fermi gases
at Rice University \cite{randy1,randy2,randy3}, MIT \cite{mit1,mit2,mit3,mit4}
and ENS \cite{ens} have stimulated a flurry of theoretical activity.
These efforts represent an important area within the general goal
of emulating complicated many-body systems using cold atoms.

All cold atom experiments are necessarily performed in the presence
of trapping potentials that hold the atoms together in an inhomogeneous
environment. In order to extract the bulk properties of the system
(e.g, the equations of state) from measurement on a trapped sample,
a local density approximation (LDA) is employed to account for the
effects of the trapping potential. The LDA states that the system
can be treated locally as a part of an infinite system. In other words,
on a length scale which is short in comparison to the spatial variation
of the external potential, the external potential $V_{{\rm ext}}({\bf r})$
essentially acts as an offset to the chemical potential. In practice,
one defines a local chemical potential $\mu({\bf r})=\mu-V_{{\rm ext}}({\bf r})$,
with $\mu$ being the global chemical potential. In the context of
cold atoms, the LDA is often an accurate approximation when there
are no phase boundaries present. However, because of the large change
in the density which occurs from the center to the edge of the trap,
there is often more than one phase present and care needs to be taken
in application of the LDA. This is especially true at phase boundaries
where the LDA usually fails. Even in the circumstances
where there is only one phase present in the trap, corrections to
the LDA become crucial when the number of particles in the sample
is small and finite-size effects are significant \cite{son}.

The trapped polarized Fermi gas represents just such a system ---
in all experimental observations, phase separation between a region
with vanishing spin population and one with finite spin population
has been observed. The purpose of this paper is to examine the validity
of the LDA as the particle number and aspect ratio of the trap is
changed. To this end we employ the Bogoliubov-de Gennes (BdG) equations
\cite{gennes}, which is a powerful mean-field tool particularly
suitable for inhomogeneous Fermi superfluids, and has been recently
adopted by many to study trapped ultra-cold Fermi gases \cite{liu3D,machida,gri,torma,Ueda_tezuka}.

In a recent paper \cite{fflo} we analyzed discrepancies observed
in pioneering experiments on polarized fermionic superfluids \cite{mit1,mit2,randy1,randy2}
which appear to center on physics arising when the containing trap
is highly elongated. In our analysis we found that as the trap becomes
increasingly elongated, the solutions of the BdG show a tendency towards
metastable behavior which could lead to the observation of
states which are not necessarily the ground state. However we also
observed that one class of solutions with structure similar to the
LDA solution was consistently the lowest in energy within our 
analysis. Our conclusions have since been confirmed by similar calculations \cite{Pei}
using a density functional formulation \cite{bulgac} which is more sophisticated
than the BdG and accounts for quantum fluctuations and interactions
within the normal fluid. In this paper we focus on \textit{this} class
of solutions and examine how well such solutions match those obtained
from the LDA approximation as a function of particle number and trap
aspect ratio. 

The content of the paper is organized as follows. In Sec.~\ref{sec:BdG-formulation-and},
we present the BdG formulation and describe the numerical techniques
used to solve the BdG equations. In Sec.~\ref{N200} we discuss our
results for a relatively small number of particles and for different
trapping geometry. In Sec.~\ref{Nbig}, we focus on a very elongated
cigar-like trap but vary the number of atoms. Finally, a concluding
remark is presented in Sec.~\ref{sum}.

\section{The Bogoliubov De-Gennes treatment \label{sec:BdG-formulation-and}}

\subsection{Formulation}

We consider a gas of spin-polarized fermionic atoms interacting through
a contact potential ($g\sum_{i<j}\delta^{3}(\vec{r}_{i}-\vec{r}_{j})$)
and confined to a harmonic trap defined in cylindrical coordinates
$(\rho,\phi,z)$ by $V_{\rm ext}(\rho,z)=\frac{m}{2}(\omega_{\bot}^{2}\rho^{2}+\omega_{z}^{2}z^{2})$
with axial and radial frequencies denoted by ($\omega_{z},\omega_{\bot})$.
We work at unitarity $(a_{s}\rightarrow\infty)$ and within a cigar-shaped
trap with aspect ratio $\alpha=\omega_{\bot}/\omega_{z}$. This system
of $N=N_{\uparrow}+N_{\downarrow}$ atoms interacting through contact
interaction is described by a Hamiltonian $\hat{H}=\int d\vec{R} \,(H_{0}+H_{I})$
with non-interacting $H_{0}$ and interacting $H_{I}$ portions given
by:
\begin{eqnarray}
H_{0}(\vec{R}) &= & \sum_{\sigma}\psi_{\sigma}^{\dagger}({-\frac{\hbar^{2}}{2m}\nabla^{2}+V_{\rm ext}
\left(\rho,z\right)-\mu_{\sigma}})\psi_{\sigma} \,,\nonumber \\
H_{I}(\vec{R}) &= & -g\, \psi_{\uparrow}^{\dagger}(\vec{R})\psi_{\downarrow}^{\dagger}(\vec{R})\psi_{\downarrow}(\vec{R})\psi_{\uparrow}(\vec{R}) \, ,\label{eq:basic_hamiltonian}\end{eqnarray}
where $\psi_{\sigma}(\vec{R})$ represents the fermionic field operators,
$m$ the mass and $\mu_{\sigma}$ the chemical potential of atomic
species with spin $\sigma$. The coupling constant is defined as $g=\frac{4\pi\hbar^{2}a}{m}$.
Henceforth, we work in trap units for which: $m=\omega_{z}=\hbar=k_{B}=1$.
This implies that energies will be measured in units of $\hbar\omega_{z}$,
lengths in units of $l_{0}=\sqrt{\frac{\hbar}{m\omega_{z}}}$ and
temperature ($T$) in units of $\hbar\omega_{z}/k_{B}$ .

The Hamiltonian (\ref{eq:basic_hamiltonian}) will be treated within
the mean-field BdG approximation for which there are many excellent
references \cite{gennes,Castin_bcs_theory,hu1}. Here we simply state
the BdG equations for the pair wave functions $u_{j}(\vec{R})$ and
$v_{j}(\vec{R})$ which decouple $\hat{H}$:

\begin{equation}
\left[\begin{array}{cc}
{\cal H}_{\uparrow}^{s}-\mu_{\uparrow} & \Delta(\vec{R})\\
\Delta(\vec{R}) & -{\cal H}_{\downarrow}^{s}+\mu_{\downarrow}\end{array}\right]\left[\begin{array}{c}
u_{j}\\
v_{j}\end{array}\right]=E_{j}\left[\begin{array}{c}
u_{j}\\
v_{j}\end{array}\right],\label{mean_field_ham}\end{equation}
In the above coupled set of equations, $u_{j}(\vec{R})$ and $v_{j}(\vec{R})$
are two components of the quasi-particle wavefunction associated with
energy $E_{j}$. The single particle Hamiltonian ${\cal H}_{\sigma}^{s}$
is defined in our trap units by:\begin{equation}
{\cal H}_{\sigma}^{s}(\vec{R})=-{\nabla^2}/{2}+V_{\rm ext}+g\rho_{\bar{\sigma}}-\mu_{\sigma}\label{eq:diagonal_h}\end{equation}
and includes the trapping potential, the chemical potential $\mu_{\sigma}$
and the Hartree mean-field potential is given by density $\rho_{\sigma}(\vec{R})=\langle\Psi_{\sigma}^{\dagger} (\vec{R} )\Psi_{\sigma}(\vec{R} )\rangle$.
In accordance with fermionic commutation relations~\cite{gennes},
the quasi-particle amplitudes are normalized as:\begin{eqnarray}
\int d\vec{R} \,|u_{j}(\vec{R})|^{2}+|v_{j}(\vec{R})|^{2}=1\label{eq:uv_normalization}\end{eqnarray}
and are related to the spin densities through :\begin{eqnarray}
\rho_{\uparrow}(\vec{R}) & = & \sum_{j=1}^{\infty}|u_j(\vec{R})|^{2}f\:(E_{j})\nonumber \\
\rho_{\downarrow}(\vec{R}) & = & \sum_{j=1}^{\infty}|v_j(\vec{R})|^{2}f\:(E_{j}),\label{eq:hartree_h}\end{eqnarray}
where $f(E)=1/(e^{E/k_{B}T}+1)$ is the Fermi-Dirac distribution function.
In the unitary limit the Hartree terms $g\rho_{\bar{\sigma}}$ on the
diagonal of Eq.~(\ref{mean_field_ham}) do not actually diverge but are
unitarity limited \cite{gupta}. How to incorporating the Hartree term in the unitarity limit is beyond the mean-field BdG formalism. For this reason we ignore this term
in our calculations. The paring field or gap paramter is give by 
\begin{equation} 
\Delta(\vec{R}) = g \langle \psi_\uparrow (\vec{R}) \psi_\downarrow (\vec{R}) \rangle = g \sum_j u_j(\vec{R}) v^*_j(\vec{R}) f(E_j) \label{eq:uv_gap_exp}
\end{equation}
Since the Hartree terms are ignored in our analysis,
Eqs.~(\ref{mean_field_ham}), (\ref{eq:uv_normalization}) and (\ref{eq:uv_gap_exp}) 
constitute a closed set of nonlinear equations which we solve self-consitently.
However as presented above, our formulation has one problem which
arises from a nasty side effect of the contact interaction. The contact
interaction assumes wrongly that all states are scattered in the same
way regardless of their incoming energy and consequently sums in contributions
from collisions at arbitrarily high energy which creates an ultra-violet
divergence. Hence the gap equation \ref{eq:uv_gap_exp} needs to be properly regularized as we now discuss.

\subsection{Regularizing the BdG equations}
Due to the assumption of contact interaction, the gap $\Delta$ is a function of the center-of-mass coordinate of the pair, $\vec{R}$. To discuss the regularization, it is more convenient to re-introduce back the relative coordinate $\vec{r}$, with which the gap is defined as \[ \Delta (\vec{R},\vec{r}) =  \langle\psi_{\uparrow}(\vec{R}+{\vec{r}}/{2})\psi_\downarrow (\vec{R}-{\vec{r}}/{2} )\rangle \]which diverges as $\frac{\Delta(\vec{R})}{2\pi r}$ when $r \rightarrow 0$ \cite{Castin_bcs_theory}.
To regularize Eq. (\ref{eq:uv_gap_exp}), one simply subtracts off the
$1/r$ divergence to obtain the regularized equation \cite{Castin_bcs_theory}:
\begin{equation}
\frac{\Delta(\vec{R})}{g}=\sum_j \, u_{j}(\vec{R})v_{j}^{*}(\vec{R})\,f(E_{j})-\frac{\Delta}{2\pi r}\label{eq:regularized_gap_equation}\end{equation}
In practice, the convergene of the sum above is quite slow and we
discuss here a numerically efficient way of evaluating $\Delta(\vec{R})$
to sufficient accuracy without undue effort. First an energy cutoff $E_{c}$
is used to break the sum of Eq.~(\ref{eq:uv_gap_exp}) into two pieces
as ff.\begin{equation}
\frac{\Delta(\vec{R})}{g}=\sum_{E_{j}<E_{c}}\, u_{j}(\vec{R})v_{j}^{*}(\vec{R})f\:(E_{j})+\frac{\Delta_{c}(\vec{R})}{g}-\frac{\Delta(\vec{R})}{2\pi r}\label{eq:cutoff_equation}\end{equation}
 The second term $\Delta_{c}(\vec{R})$ is an approximation to the sum above
the cutoff using the LDA result for the pairing field \cite{hu1}
which can also be written as :\begin{eqnarray}
\frac{\Delta_{c}(\vec{R})}{g} & = & \frac{\Delta(\vec{R})}{(2\pi)^{3}}\int_{k_c}^{\infty}\frac{d^{3}k}{\sqrt{(\frac{k^{2}}{2}-\mu(\vec{R}))^{2}+\Delta^{2}}}
\label{eq:lda_unregularized_gap}\end{eqnarray}
where $k_c$ is the momentum cutoff related to $E_c$. This leads to a computationaly efficient form of the gap equation: 
\begin{equation}
\frac{\Delta(\vec{R})}{U_{\rm eff}(\vec{R})}=\sum_{E_{j}<E_{c}}\, u_{j}(\vec{R})v_{j}^{*}(\vec{R}) \,f(E_{j}).\label{eq:numerical_gap}\end{equation}
Here we have employed the identity: \begin{equation}
\int_{0}^{\infty}\frac{d\vec{k}}{(2\pi)^{3}}\frac{e^{ikr}}{\frac{k^{2}}{2}}=\frac{1}{2\pi r}\label{eq:divergence_integral}\end{equation}
 to subsume the LDA approximation of the gap ($\Delta_{c}$) into
an effective interaction defined by:\begin{equation}
\frac{1}{U_{\rm eff}(\vec{R})}=\left[\frac{1}{g}-\int_{k_c}^{\infty}d^{3}k
\left(\frac{1}{\sqrt{(\frac{k^{2}}{2}-\mu(\vec{R}))^{2}+\Delta^{2}}}-\frac{1}{\frac{k^{2}}{2}}\right)\right]
\label{eq:effective_interaction}\end{equation}
Below this cutoff $E_{c}$, the quasiparticle states are calculated
exactly by solving Eqs.~(\ref{mean_field_ham}), (\ref{eq:numerical_gap})
and (\ref{eq:effective_interaction}) self-consistently along with the
normalization conditions:\begin{eqnarray}
N_{\sigma} & = & \int \, \rho_{\sigma}(\vec{R}) \,d\vec{R}\label{normalization}\end{eqnarray}
which conserve total particle number $N=N_{\uparrow}+N_{\downarrow}$ and
overall polarization $P=(N_{\uparrow}-N_{\downarrow})/N$. The iterative
solution of these equations is achieved using a modified Broyden's
approach \cite{Johnson} which is a nonlinear mixing scheme allowing
the formation of polarized regions even if they were not present in
the initial condition. In this scheme convergence was achieved when
the root mean squared difference between $\Delta$ at different iterations
was below some tolerance i.e., $\sqrt{\frac{\sum_{j}(\Delta_{i}^{j}-\Delta_{i+1}^{j})^{2}}{\sum\Delta_{i}^{j2}}}<tol$,
where $j$ is the position index and $i$ represents the iteration
number. We should point out that at unitarity this method is analogous
to a descent technique where the step is optimized through the residuals
stored from a few previous steps. As mentioned previouly we choose
as our initial condition $\Delta_{\rm LDA}$, the LDA solution to the
BdG equations. This is an important point becasue we found in previous
work \cite{fflo} that at larger particle numbers the solution of
the BdG can be quite sensitive to the initial conditions.

\subsection{Special features}

We discretize using a linear triangular finite element mesh in the
$\rho$-$z$ plane which anticipates that our results will retain
the cylindrical symmetry of the confining potential $V_{ext}$. The
accuracy of these calculations are controlled by the density of the
trianglular mesh and the cut-off $E_{c}$ used in the hybrid scheme.
Both of these are changed in successive solutions until the free-energy
or relevant observable converges to a sufficient accuracy. Experience
has taught us that this simple renormalization scheme typically converges
when the cutoff is of the order $6E_{F}$ (where $E_{F}$ is the Fermi
energy) which implies that the number of quasiparticle states to be
directly calculated by Eq.~(\ref{mean_field_ham}) is about $6N$. Note
that this puts a constraint on the density of the discretizing mesh.
Thus, for moderate system sizes, we are still presented with a very
large problem. For example, for $N\sim10^{3}$ particles, one essentially
needs to calculate $\sim 10^{5}$ quasiparticle states at each iteration.

One important consequence of our finite element discretization is
that it yields sparse matrices which are suitable to massively parallel
matrix computations. This is of key importance given that the slow
convergence of the sum in Eq.~(\ref{eq:numerical_gap}) condemns us to calculate
a very large number of quasi-particle states. This is true inspite
of our efficient hybrid scheme, without which calculation would be
prohibitive. It is immediately obvious that these difficulties will
increase with the number of particles $N$, and will make the problem
impractical for even moderate particle numbers without very careful
formulation. In our case these difficulties are inescapable since
the issues to be addressed occur in the presence of finite size effects
and confinement. Hence it was crucial to develop the ability to perform
calculations with realistic particle numbers because it is not {\em
a priori} obvious how physical properties will scale with system
size. At each iteration, we need to find a large number of eigenvalues
and eigenfunctions for large matrices. To this end, we use a novel
shift-and-invert scheme which we developed independently but is very
similar to the one described in Ref.~\cite{zhang}. 

Briefly, the scheme involves partitioning the sought spectrum amongst
groups of processors working independently. The size of the group
is determined by the minimum number of processors with enough total
memory to store the inverted matrix which is required for building
the local Krylov basis. The main challenge here is bookkeeping to
prevent over-counting of states and balancing the load amongst the
processor groups. It is conceivable that this method could have issues
in cases where the equations support huge degenerate subspaces. In
our particular formulation we exploited cylindrical symmetry and parity
along the long trap axis to reduce the problem. Consequently we only
had to contend with accidental degeneracies. A good analysis of these
issues may be found in Ref.~\cite{zhang} but a more thorough description
of the numerical details will be presented elsewhere. However we note here that
this parallelization scheme is very efficient on distriubted computing
systems and scales easily to thousands of CPU's which is as high as
we have tested. Potentially it can be used to study much larger systems
than we have reported here or in \cite{fflo}.

\subsection{Validity of the BdG}

We should devote a few lines here to comment on the validity of the
BdG theory at unitarity becasue it is expected that quantum fluctuations
and other effects due to the strong interactions could be significant
in this regime. The main drawbacks of the BdG is that it fails to
account for phase fluctations. At unitarity it has an additional disadvantage
in that it also fails to account for interactions within the normal
fluid which is unitarily limited~\cite{gupta}. However the BdG is
widely expected to yield qualitatively reliable answers for two main
reasons. First because of the finite size of these experiments, the
trapped gas enjoys protection from fluctuations of arbitrarily low
energy or of very long wavelengths. Secondly due to experimental evidence
for superfluidity at unitarity, it is quite clear that interactions
within the normal fluid are not so great that the order parameter
cannot form or will be destroyed. Thus the failure to account for
these effects is not expected to change the topology of the phase
diagram but at most would slightly shift the phase boundaries. Since
our purpose to examine the suitability of the LDA is qualitative,
we are confident that the BdG can account for the essential physics. Nevertheless, due to the limitations within the BdG formalism, in particular the neglect of interaction in the normal fluid, our calculation fails to quantitatively locate position of the Clogston limit. In addition, it cannot be applied to study a system with extremely large polarization (i.e., $P \approx 1$) where the polaron physics will dominate \cite{ens,pol}.

\section{Results for $N=200$}

\label{N200} In this section, we focus on a relatively small particle
number of $N=200$. As we will show, the system is rather sensitive
to the trap geometry. In the following, we first briefly discuss the
case of a spherical trap with $\alpha=1$ and then concentrate on
elongated cigar-like trapping potentials with $\alpha>1$ and then
discuss them in detail.

\subsection{Spherical trap}

Liu {\em et al.} studied a spin-imbalanced Fermi gas confined in
a spherical harmonic trap in Ref.~\cite{liu3D}. To benchmark our
work, we first did a series of calculations for this geometry and
found our results in perfect agreement with those reported in Ref.~\cite{liu3D}.
In this case, even though we anticipate only cylindrical geometry,
the density profiles always obey the spherical symmetry. Note that
the authors of Ref.~\cite{liu3D} solved the one-dimensional (1D)
radial equation, hence the spherical symmetry of the cloud is automatically
imposed. We refer the readers to Ref.~\cite{liu3D} for details;
here we give just a brief description of the key features. The density
profiles indicate a phase-separation scenario: a fully paired BCS
superfluid core at the trap center surrounded by a fully polarized
shell composed of excess majority spins. A thin layer of partially
polarized gas forms the interface between the superfluid core and
the normal shell. In this intermediate regime, the minority density
and the order parameter sharply drop to zero. Here and in other cases,
we always found that the profile of the order parameter closely follows
that of the density of the minority spin component. Furthermore, in
this case, the LDA gives very good agreement with the full BdG calculation
even for particle numbers as small as a few hundred.

\subsection{Cigar trap}

\begin{figure}
\begin{centering}
\includegraphics[width=9cm]{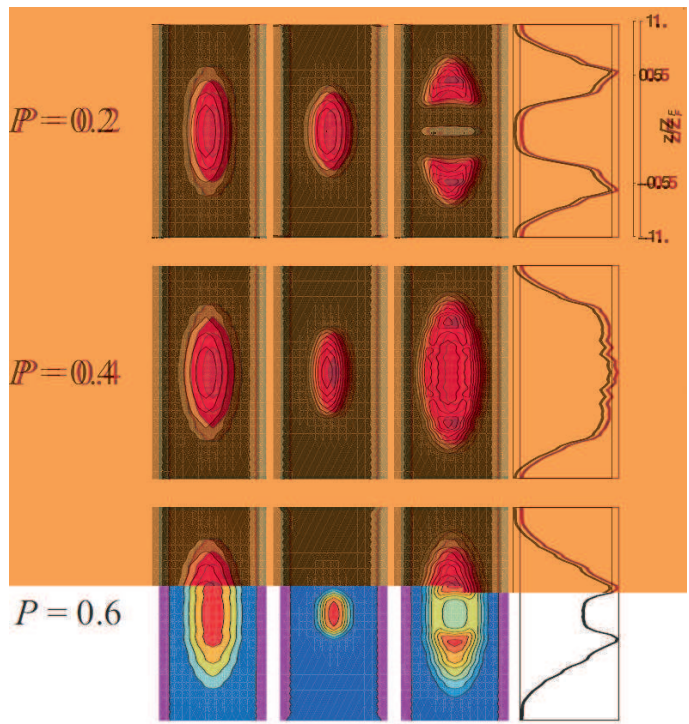} 
\par\end{centering}

\caption{\label{ar5c} Density profile of the atomic clound for $N=200$ in
an elongated trap with aspect ratio $\alpha=5$. The upper, middle
and lower row are results from different polarizations $P=0.2$, 0.4
and 0.6, respectively. In each row, we have shown (from left to right)
the column densities of the majority component $\int dx\,\rho_{\uparrow}$,
the minority component $\int dx\,\rho_{\downarrow}$, their difference
$\int dx\,(\rho_\uparrow-\rho_{\downarrow})$, and the axial spin density
$n_{1D}(z)=\int dxdy\,(\rho_{\uparrow}-\rho_{\downarrow})$.}

\end{figure}
\begin{figure}
\begin{centering}
\includegraphics[width=10cm]{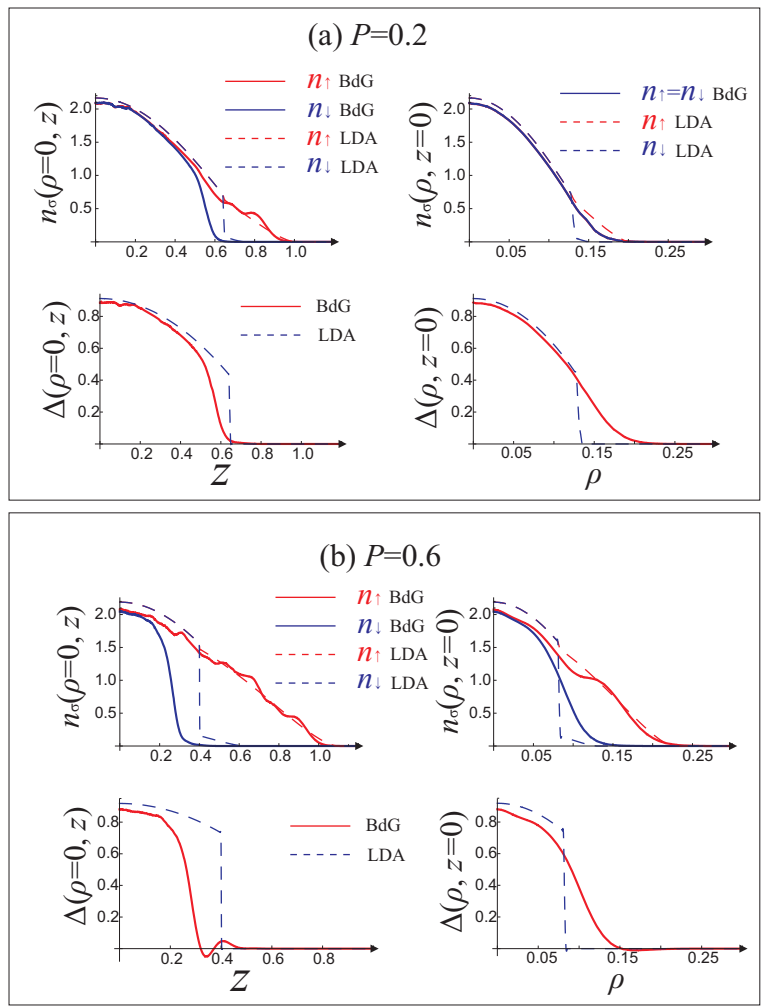} 
\par\end{centering}

\caption{Density and order parameter profiles along the axial and radial axes in a cigar-like trap with $\alpha=5$ for two differnet polarizations: (a) $P=0.2$ and (b)
$P=0.6$. }

\label{fig2} 
\end{figure}

So far, all the experiments on spin-imbalanced Fermi gases have been
performed in cigar-like traps with aspect ratio $\alpha>1$. For a
given particle number, the 1D regime is eventually encountered in
this geometry by increasing $\alpha$ and creates the possibility
to study the 3D-1D dimensional crossover. Figure~\ref{ar5c} illustrates
several examples of the density profiles for $N=200$ atoms confined in a moderately elongated trap with $\alpha=5$
(this trap aspect ratio is close to what has been used in the MIT
experiments). We find it convenient to express our results in terms
of the Fermi energy $E_{\textit{F}}=(3N)^{1/3}\alpha^{2/3}$, central
number density $(2E_{F})^{3/2}/(6\pi^{2})$, and the Thomas-Fermi
radius along the $z$-axis $Z_{\textit{F}}=\sqrt{2E_{\textit{F}}}$
for a single species ideal Fermi gas of $N/2$ particles in a trap
with identical parameters. The upper row of Fig.~\ref{ar5c} shows
the density profiles of a system with a relatively small polarization
$P=0.2$. Here the axial spin density $n_{1D}(z)$ exhibits a double-horn
structure and vanishes near $z=0$. This is a clear violation of the
LDA which predicts that $n_{1D}$ should be flat topped \cite{LDAe}.
Fig.~\ref{ar5c} can be examined in tandem with Fig.~\ref{fig2} where
for a closer inspection, we plot the densities and the order parameter
along the axial and radial axis for two different polarizations. Fig.~\ref{fig2}(a)
displays results for $P=0.2$.
The density profiles along the $z$-axis show clearly a phase separated
three-region structure --- moving from the center to the edge of the
trap, we encounter a fully paired superfluid core, a partially paired
intermediate region and a fully polarized normal gas, just like in
the previous case of spherical trap. In stark contrast, the density
profiles for the two components along the $\rho$-axis are completely
overlapped. In fact, this matching of the radial profiles occur for
$|z|\le0.1$. As a consequence, the axial spin density vanishes near
$z=0$ as shown in the upper row of Fig.~\ref{ar5c}.

That the majority and minority densities overlap along the radial
direction can be understood from an argument invoking the surface
energy. When induced phase separation occurs, there is an accompanying
surface energy associated with the interface between the two phases.
The system will then try to minimize the interface in order to reduce
the associated energy. For a cigar-like trap as we study here, the
superfluid-normal gas interface area can be efficiently reduced if
the two spin components match their densities radially. The authors
of Ref.~\cite{STmueller,STStoof} devised phenomenological theories
to include the surface term variationally to explain the breakdown
of the LDA observed in the Rice experiment~\cite{randy1,randy2}.
In our calculation, the surface energy is automatically included from
the self-consistent BdG formulation \cite{adilet}.

As polarization increases, eventually it becomes energetically unfavorable
to have this radial overlap. This is illustrated in Fig.~\ref{fig2}(b)
for $P=0.6$. Consequently, the axial spin density no longer vanishes
near $z=0$ and the LDA becomes more accurate (see the middle and
bottom rows of Fig.~\ref{ar5c}). In addition, it is quite noticeable
that, particularly for large $P$, the minority component density
has a steeper down turn along the axial axis than along the radial
axis. Moreover, in the partially polarized intermediate region, the
order parameter has a small oscillation along the axial axis, but
not along the radial axis. Similar order parameter oscillations were
also found in the spherical trap case \cite{liu3D}. This is a consequence
of the proximity effect which, in the context of superconductor, occurs when a superconductor is in contact with a normal metal, the Cooper pairs from the superconductor diffuse into the normal component.

\begin{figure}
\begin{centering}
\includegraphics[width=10cm]{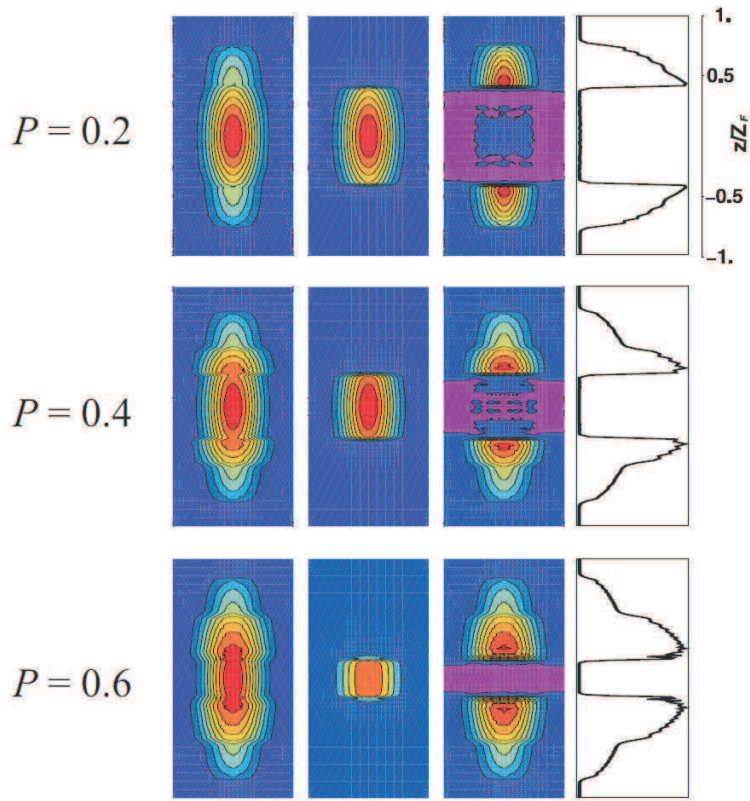} 
\par\end{centering}
\caption{\label{ar50c} Same as in Fig.~\ref{ar5c}, but for $\alpha=50$.}
\end{figure}

\begin{figure}
\includegraphics[scale=0.75]{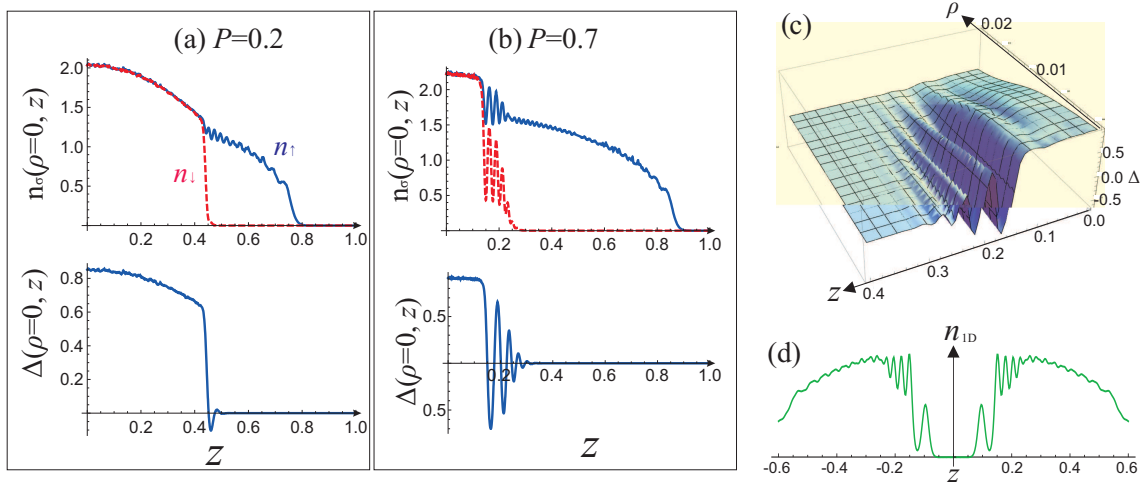}
\caption{$\alpha=50$ for two different polarizations: (a) $P=0.2$ and (b)
$P=0.7$. Same units as in Fig.~\ref{fig2}. We do not show the radial
density and order parameter profiles, which look more or less like
those in Fig.~\ref{fig2}(a). (c) The surface plot of the order parameter
for the case of $P=0.7$ in the $\rho$-$z$ plane, showing strong
oscillations with the nodes aligned along the radial direction. (d)
The density oscillations for $P=0.7$ leave a strong signal in the
doubly integrated axial spin density $n_{1D}$.}
\label{fig4} 
\end{figure}

Next, we keep $N$ fixed at 200 but increase the trap aspect ratio
to $\alpha=50$ which represents a much more elongated cigar trap
and close to what is used in the Rice experiment. A similar display
of the column and axial spin density profiles for different polarization
as in Fig.~\ref{ar5c} is shown in Fig.~\ref{ar50c}. In this very
elongated trap, the majority and minority components have their densities
matched along the radial axis up to the highest polarization we have
calculated which is $P=0.7$, and the minority component has a boxy-looking
density profile. This further confirms that the system is able to
greatly reduce the effective surface area between the normal state
and the superfluid state in anisotropic cigar-like traps. Another
marked feature for such an elongated trap is the prominent oscillations
of the order parameter along the $z$-axis. As demonstrated in Fig.~\ref{fig4},
these oscillations are quite generic features in such a trap with
finite $P$. As $P$ increases, both the amplitude and the spatial
extension of the oscillations increase. As shown in Fig.~\ref{fig4}(b),
at large polarizations, the axial length of the partially polarized
intermediate region becomes comparable to or even larger than that
of the BCS core. Accompanied by the oscillation in the order parameter,
the density profiles (in particular, the minority density) also exhibit strong oscillations. Such oscillations are reminiscent of the FFLO pairing state
predicted by Fulde, Ferrel, Larkin and Ovchinnikov \cite{FF_original,LO_original}
in which the Cooper pairs possess finite momentum and the order parameter
in the bulk develops sinusoidal oscillations that break the spatial
translation symmetry.

Our calculations also show that these axial oscillations are aligned
along the radial axis, as shown in Fig.~\ref{fig4}(c). We even intentionally
started from an initial ansatz of $\Delta$ where the axial oscillations
are present but with the nodes mis-aligned in the radial direction,
the BdG iterations eventually converge to a state where the nodes
are perfectly aligned radially. This radial alignment has important
impact in detecting the oscillations in column density profiles where
the densities are integrated along one radial axis: Due to the radial
alignment, the oscillations are not washed out and can be easily observed,
for example, in the doubly integrated axial spin density $n_{1D}$,
as illustrated in Fig.~\ref{fig4}(d).

It is interesting to compare our result with the recent work by Bulgac
and Forbes \cite{bulgac} who, using a density functional theory,
argued that the FFLO pairing phase occupies a larger phase space region
than people previously thought for a 3D homogeneous system. The FFLO
state found in Ref.~\cite{bulgac} is also associated with large-amplitude
density oscillations, particularly in the minority component. Another
perspective on the order parameter oscillations and its potential
connection with the FFLO phase is dimensionality. It is well known
that the partially polarized phase with FFLO-like oscillations is
prominently featured in the phase space of 1D systems \cite{hu1,orso,hu}.
That the reduced dimensionality favors such a state can be understood
from the argument of Fermi surface nesting or alternatively from the
cost of creating domain walls. The cigar-like traps used in our calculation
mimic a quasi-1D system and may be the reason that we see pronounced
oscillations in our calculation. If this latter explanation is correct,
i.e., the partially polarized region featuring FFLO-like oscillations
is due to the effective reduction of the spatial dimension, we then
expect to see these oscillations diminish as $N$ is increased while
the trap aspect ratio is fixed, which makes the system more 3D-like.
To confirm this, we now turn to the next section where we keep $\alpha=50$
but vary the total particle number $N$.

\section{Results for large particle numbers at $\alpha=50$}

\label{Nbig}

\begin{figure}
\begin{centering}
\includegraphics[width=10cm]{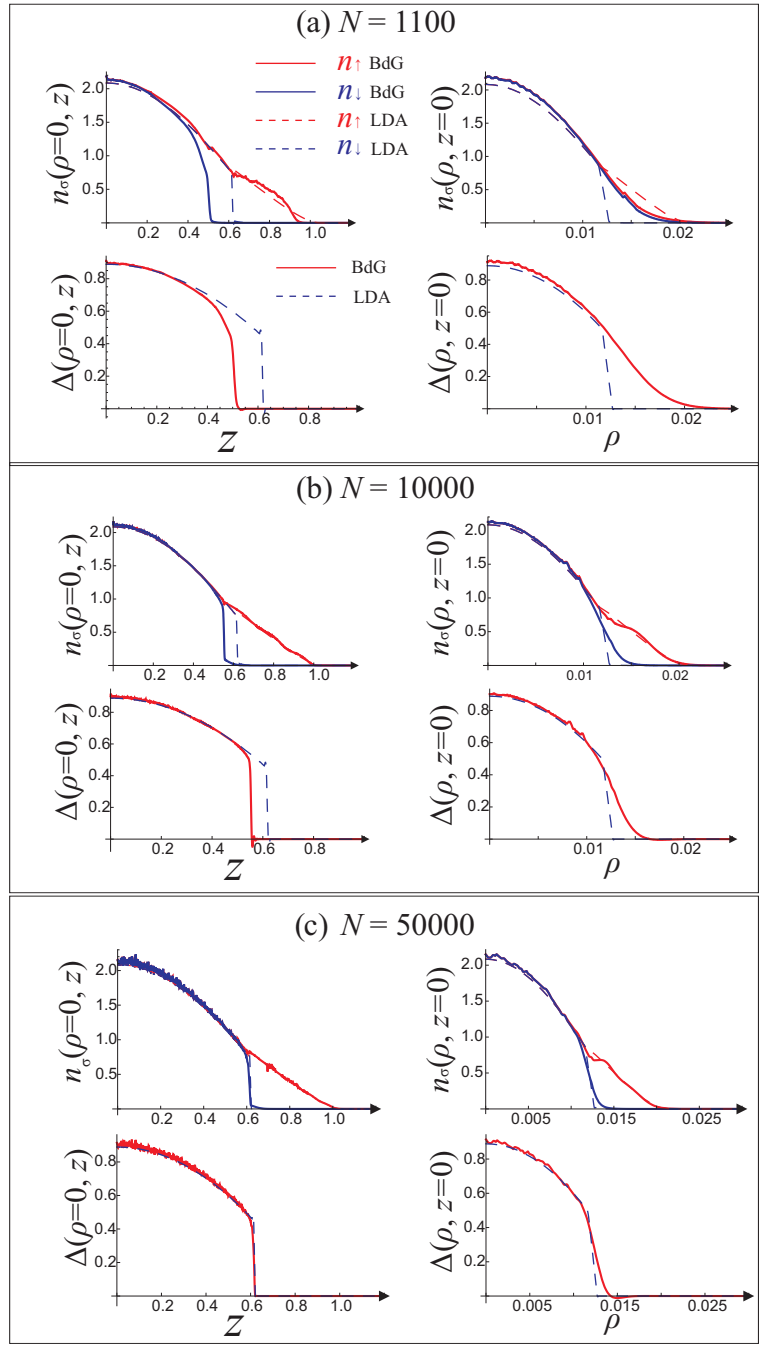} 
\par\end{centering}
\caption{Density and order parameter profiles along the axial and the radial axis in a cigar-like trap with $\alpha=50$ for $P=0.3$ but different values of total particle number
$N$. Same units as in Fig.~\ref{fig2}.}
\label{fig5} 
\end{figure}

We now consider a trapped system in a very elongated cigar trap with
$\alpha=50$. Fig.~\ref{fig5} shows the density and order parameter
along the axial and radial axis at a fixed polarization $P=0.3$
but different values of total particle number $N$. As one can clearly
see, the oscillations in both the order parameter and the density
profiles diminish as $N$ is increased and the LDA approximation becomes
more and more accurate, which indicates that the FFLO-like region
observed above for small $N$ does not represent a bulk 3D phase.
Rather, it is a finite-size effect due to the effective reduction
of the spatial dimension.

Nevertheless, we have discovered that as $N$ increases, the system
exhibits a tendency towards metastability \cite{fflo}. Numerically,
by starting from different initial ansätze for the order parameter
$\Delta(\vec{r})$, the BdG solution may converge to different final
states. Our calculations show that among these different states, the
one that closely resembles the LDA solution always has the lowest
energy as long as $N$ is sufficiently large ($N \gtrsim 10^{4}$),
but that there may exist several metastable states with energies just
slightly larger that violate the LDA. The observed LDA-violating states
at Rice are most likely these metastable states. Experimentally, whether
the ground state or a metastable state will be realized may depend
upon how the evaporative cooling procedure is implemented \cite{Parish_transport}.
This has been confirmed very recently in a new experiment by Hulet
group \cite{hulet}.

\section{Conclusion}

\label{sum}

In conclusion, we have carried out a systematic study of a trapped
spin-imbalanced Fermi gas in the unitary limit up to a total number
$N\sim10^{5}$ atoms. We study a class of solutions which has recently
been identified as having the lowest energy \cite{fflo,Pei} through
a self-consistently solution of the Bogoliubov-de Gennes equations
using state-of-the-art numerical techniques. For a given set of trapping
parameters, the LDA will eventually become accurate for sufficiently
large $N$. However, the validity of the LDA is also sensitive to
the trap geometry: Traps with small anisotropy favor the LDA. Our calculations show that for
a relatively small number of atoms in a very elongated trapping potential,
the system contains three phases: an unpolarized BCS phase, a partially
polarized FFLO-like phase and a normal phase. That the FFLO region
exists may be understood from the view point of reduced effective
spatial dimension. As $N$ is increased while all other parameters
remain fixed, the FFLO-like region eventually disappears. A detailed
analysis of dimensional crossover will be reported in the future.

\ack This work was supported by a grant from the ARO with funding from
the DARPA OLE program, the Welch foundation (C-1669, C-1681) and NSF.
We thank the hospitality of KITP where part of the work is carried
out.

\end{document}